\documentclass{aastex631}
\usepackage{booktabs}
\usepackage{savesym}
\savesymbol{tablenum}
\usepackage{siunitx}
\usepackage{multirow}
\usepackage{longtable}
\usepackage{graphicx}
\usepackage{supertabular}
\usepackage{tabularx}
\shorttitle{Fine-grained CNN Model for Solar Flare Forecasting}
\shortauthors{Deng et al.}
\graphicspath{{./}{figures/}}

\begin{document}

\title{Fine-grained Solar Flare Forecasting Based on the Hybrid Convolutional Neural Networks\footnote{Released on March, 1st, 2021}}

\correspondingauthor{Feng Wang and Hui Deng}
\email{fengwang@gzhu.edu.cn,denghui@gzhu.edu.cn}

\author[0000-0003-1485-4608]{Zheng Deng}
\affiliation{Center For Astrophysics, Guangzhou University, Guangzhou 510006, P.R. China}
\affiliation{Faculty of Information Engineering and Automation, Kunming University of Science and
Technology, Kunming 650500, P.R. China}
\affiliation{Great Bay Center, National Astronomical Data Center, Guangzhou, Guangdong, 510006, P.R. China}

\author[0000-0002-9847-7805]{Feng Wang}
\affiliation{Center For Astrophysics, Guangzhou University, Guangzhou 510006, P.R. China}
\affiliation{Great Bay Center, National Astronomical Data Center, Guangzhou, Guangdong, 510006, P.R. China}

\author[0000-0002-8765-3906]{Hui Deng}
\affiliation{Center For Astrophysics, Guangzhou University, Guangzhou 510006, P.R. China}
\affiliation{Great Bay Center, National Astronomical Data Center, Guangzhou, Guangdong, 510006, P.R. China}

\author{Lei. Tan}
\affiliation{Center For Astrophysics, Guangzhou University, Guangzhou 510006, P.R. China}

\author{Linhua Deng}
\affiliation{Yunnan Observatories, Chinese Academy of Sciences, Kunming 650216, P.R. China}

\author{Song Feng}
\affiliation{Faculty of Information Engineering and Automation, Kunming University of Science and
Technology, Kunming 650500, P.R. China}



\begin{abstract}

Improving the performance of solar flare forecasting is a hot topic in solar physics research field. Deep learning has been considered a promising approach to perform solar flare forecasting in recent years. We first used the Generative Adversarial Networks (GAN) technique augmenting sample data to balance samples with different flare classes. We then proposed a hybrid convolutional neural network (CNN) model ($M$) for forecasting flare eruption in a solar cycle. Based on this model, we further investigated the effects of the rising and declining phases for flare forecasting. Two CNN models, i.e., $M_{rp}$ and $M_{dp}$, were presented to forecast solar flare eruptions in the rising phase and declining phase of solar cycle 24, respectively. A series of testing results proved: 1) Sample balance is critical for the stability of the CNN model. The augmented data generated by GAN effectively improved the stability of the forecast model. 2) For C-class, M-class, and X-class flare forecasting using Solar Dynamics Observatory (SDO) line-of-sight (LOS) magnetograms, the means of true skill statistics (TSS) score of $M$ are 0.646, 0.653 and 0.762, which improved by 20.1\%, 22.3\%, 38.0\% compared with previous studies. 3) It is valuable to separately model the flare forecasts in the rising and declining phases of a solar cycle. Compared with model $M$, the means of TSS score for No-flare, C-class, M-class, X-class flare forecasting of the $M_{rp}$ improved by 5.9\%, 9.4\%, 17.9\% and 13.1\%, and the $M_{dp}$ improved by 1.5\%, 2.6\%, 11.5\% and 12.2\%. 

\end{abstract}




\section{Introduction} \label{sec:intro}

A solar flare is one of the most intense solar activities, manifested mainly in the sudden enhancement of the radiation flux from the radio band up to the X-rays. Flares have a significant impact on Earth's space environment, and flares also have different degrees of direct or indirect effects on meteorology and hydrology.
The study of flares, especially forecasting flare eruptions, has become a research hotspot and an essential element of space weather forecasting.

As early as the 1930s, solar physicists began to forecast solar flare eruptions. \cite{giovanelli1939relations} proposed a forecast model based on the statistical relationship between flares and sunspots. \cite{2002Active} and \cite{2012Toward} used the Poisson statistical model to estimate the probability of solar flares. \cite{leka2003photospheric} and \cite{2007Probabilistic} applied the discriminant analysis to determine the significance of magnetic parameters of solar flares. By using the superimposed epoch analysis method, \cite{mason2010testing} found the relationship between magnetic field parameters and solar flares.

With the rapid development of computer technology,  machine learning technology was applied to the forecasting of solar flares and played an important role. For examples, the support vector machine \citep{yuan2010automated,bobra2015solar,sadykov2017relationships,hazra2020distinguishing}, the artificial neural networks \citep{qahwaji2007automatic,li2013solar,nishizuka2018deep}, the Bayesian network methods \citep{yu2010short}, the random forest algorithm \citep{liu2017predicting,2018Forecasting,hazra2020distinguishing}, the ensemble learning \citep{colak2009automated}, and the K-nearest neighbor \citep{hazra2020distinguishing}. \cite{barnes2016comparison} compared different machine learning algorithms for solar flare forecasting and demonstrated the importance of making such systematic comparisons, and of using standard verification statistics to determine what constitutes a good prediction scheme.

Deep learning neural network is a hot research topic with strong learning ability and good adaptability. It gradually develops into a mainstream technique for large-scale analysis in astrophysical research. 
For example, recurrent neural network (RNN) was used by many studies because it uses sequence data for training which can mine certain information such as timing information. \cite{chen2019Identifying} adopted long short-term memory (LSTM), a special RNN, to capture both spatial and temporal information to identify the precursors of solar flare events. \cite{wang2020predicting} used LSTM to forecast solar flares and further investigated solar cycle dependence. \cite{jiao2020solar} developed a mixed LSTM regression model to forecast the maximum solar flare intensity and found that an efficient period for forecasting the solar activity is within 24-hour before the forecasting time using the LSTM.

The convolutional neural network (CNN) \citep{lecun2015deep}, a prevalent deep learning method, can automatically extract features in the field of image processing and computer vision.Compared with the classical machine learning algorithms, CNN contains several hidden layers.
Each layer extracts complex features from the data before performing a classification or regression task, CNN loses less information and is more suitable for high-dimensional data. 

Therefore, many research works have carried out flare forecasting using the line-of-sight magnetograms obtained by CNN on Solar Dynamics Observatory (SDO) observations. 
\cite{park2018application} proposed a CNN model, which is a combination of Google \citep{szegedy2015going} and Densenet \citep{2018Deep}, and used full-disk solar LOS magnetograms to make binary class predictions within 24 hr. \cite{huang2018deep} presented a model based on CNN for flare prediction via binary classification with 6, 12, 24, 48hr. \cite{zheng2019solar} proposed a hybrid CNN model based on multi-class classification for solar flare forecast. \cite{li2020predicting} further developed the ability for forecasting $\geq$ C and $\geq$ M soalr flares based on VGG net \citep{simonyan2014very} and Alex net \citep{krizhevsky2012imagenet}. Their model was built on several active regions from the LOS magnetograms, the active regions selected for modeling are located within $\pm 45^{\circ}$ of the center of the solar disc to avoid the projection effects.

These previous studies for forecasting solar eruptions using machine learning have yielded valuable results and have played an active role in promoting space weather forecasting. However, there are still some shortcomings in the current research. Firstly, the number of samples is insufficient, especially the X-class flare samples. Secondly, all the current research work has been done for the whole solar cycle, and there is a lack of finer-grained research, such as the influence of rising and falling phases on flare forecast.

In this study, we propose and validate a more stable hybrid CNN model for forecasting flares within 24 hours in a full solar cycle. Based on this model, we further investigate the performance of fine-grained models on rising and declining phases of solar cycle 24, respectively. We introduced data preparation and data argumentation using the Generative Adversarial Networks (GAN) in Section~\ref{section_data}. A stable CNN model for forecasting flare eruptions in a full cycle data is presented in Section~\ref{sec:solar_cycle_cnn_model}. In Section~\ref{sec:fine-grained_cnn_model}, we further presented two fine-grained hybrid CNN models in the rising phase and declining phase. We discuss several issues in modeling and data analysis in Section~\ref{sec:discussion}. The conclusions and future works are presented in the last Section.

\section{data preparation} \label{section_data}

\subsection{Data Source }

Helioseismic and Magnetic Imager (HMI) is one of three instruments aboard the SDO designed to study oscillations and the magnetic field at the solar surface. HMI observes the full solar disk at 6173 Å with a resolution of 1 arc second.
HMI makes two independent measurements of the LOS component of the photospheric magnetic field. One is collected every 45 seconds with the HMI Doppler camera. The other is computed every 720 seconds using filtergrams recorded by the Vector Field camera. The spatial resolution is 1 arc-second (half arc-second pixels), and the full disk images are collected on a 4096$\times$4096 detector. The noise level is nominally between 5 and 10 Gauss. 


Near the end of 2012, Space-weather HMI Active Region Patches (SHARP) data, which is convenient for AR event forecasting, were released~\citep{2014The}. \cite{zheng2019solar} used SHARP as original data and built an excellent flare forecast model. However, many flare event records of SHARP lack location and National Oceanic and Atmospheric Administration (NOAA) active region numbers. Therefore, we use the magnetograms with a cadence of 720 seconds (hmi.M\_720s) of SDO/HMI LOS magnetogram data downloaded from Joint Science Operations Center, ranging from May 2010 to November 2019. 
We also select the LOS magnetograms data located within $\pm 45^{\circ}$ of the central meridian to decrease the projection effects \citep{ahmed2013solar,2014Solar}.

Considering the data downloaded are full disk magnetograms, we collected the relevant position of active regions from \url{https://www.solarmonitor.org/index.php} and cut out the related section concerning solar flares with the size of 512$\times$512. Considering GAN and CNN require that all input images must have a fixed size, we resized all images to 128$\times$128 images just like \cite{zheng2019solar} and \cite{li2020predicting} did. 

\subsection{Data Set}

To compare with the results of the previous work, especially with the results of \cite{zheng2019solar}, we followed strictly on the method given in the study of \cite{zheng2019solar} to label the flare class with corresponding magnetograms of the active area and finally generate a data set ($D$) for the study.
The data set $D$ includes the magnetogram data labeled with four labels, i.e., No-flare, C-class, M-class, and X-class. The list of data volumes for each level of flare is shown in Table~\ref{origin_data_number}. 

\begin{table}[htp]
\centering
\caption{The Number of Solar Magnetogram Samples We Collected}
\label{origin_data_number}
\begin{tabular}{ccccc} 
\toprule
Solar Class & No-flare Class & C Class& M Class& X Class \\
\midrule
Number & 25680 & 13440 & 9120 & 1858 \\
\bottomrule
\end{tabular}
\end{table}

\subsection{GAN And Data augmentation}\label{sec:gan}

Due to the significant difference in the number of samples for each flare class, we augmented the X-class samples to balance the proportion of the sample data. In general, 
Over-sampling is an easy method for balancing training sets that repeatedly sample the minority classes with a relatively low proportion in the training set, but over-sampling does not prevent over-fitting. 

\begin{figure}
    \centering
    \includegraphics[width=18cm]{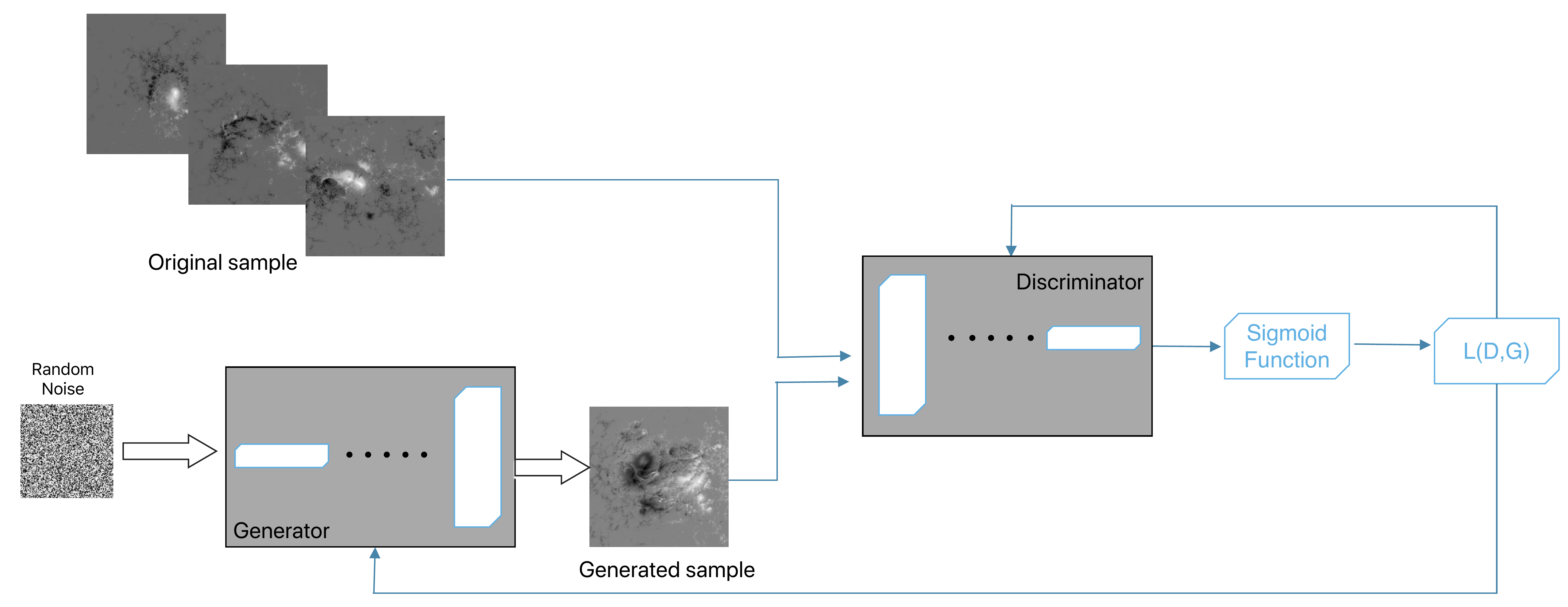}
    \caption{The training process of GAN}
    \label{fig:Gan}
\end{figure}
We chose GAN to augment the data. The GAN consists of two neural networks, one is the Generator (G) and the other is the Discriminator (D), which competes with each other over the available training data to improve the data quality of the generated \citep{goodfellow2014generative,radford2015unsupervised}. Figure~\ref{fig:Gan} shows the training process of the GAN in the study. The structures of $GNet$ and $DNet$ were designed for the GAN, respectively (see Table \ref{label:G}). 

The input of $G Net$ is a random noise array (see Table \ref{label:G}), and an image with 1$\times$256$\times$256 will be output after fully connected layers.
$GNet$ contains 6 convolutional layer, we respectively chose Batch Normalization (BN) \citep{DBLP:journals/corr/IoffeS15} and Rectified linear unit (ReLU) \citep{10.5555/3104322.3104425} as regularization and activation function for each convolutional layer to speed up the training rate of the model and get better results.

The sample images are first convoluted by the first convolutional layer with 5$\times$5 kernel size. 
The first convolutional layer follows five  convolutional layers with 3$\times$3 kernel size. 

The generated samples are further mixed with original samples and input to $D Net$. $D Net$ consists of 9 layers. The first three layers contain a convolutional layer with 5$\times$5, a max-pooling layer with 2$\times$2, and an avg-pooling layer with 2$\times$2 kernel size. 
The subsequent three layers are the same as the first three, except for the parameter of padding. The feature map obtained by convolution is output to the fully connected layer after linearization in FL7 (see Table \ref{label:G}). The activation function of the first fully connected layer is LeakyReLU \citep{maas2013rectifier} which is the same as convolutional layers. For the last activation function, we chose sigmoid \citep{bergstra2009quadratic} to output the judgment result.

Meanwhile, the data set is undersampled by randomly selecting No-flare and C-class samples to balance the data set with about 4 samples and 8 samples per 10 samples, respectively. Finally, we generate a new data set $D$ shown in Table \ref{GAN_data_number}. On the basis of $D$, we prepared 6 independent training sets and testing sets (see Table \ref{table:m}) for the subsequent modeling.

\begin{table}
\centering
\caption{The structure of $G Net$ and $D Net$}
\begin{tabular}{lcccccccc}
\toprule
Layer & Type & INPUT & Kernel Size & Striding & Padding & Regularization & Activation &  OUTPUT \\
\midrule
\midrule
IN & INPUT & 1$\times$2000 & ... & ... & ... &  ... & ... & 1$\times$65536 \\
FC1 & Fully connected & 1$\times$65536 &...&...&... & BN & ReLU & 1$\times$256$\times$256\\
CON2 & Convolution & 1$\times$256$\times$256 & 5$\times$5 & 1 & 1  & BN & ReLU & 64$\times$254$\times$254\\
CON3 & Convolution & 64$\times$254$\times$254 & 3$\times$3 & 1 & 1  & BN & ReLU & 64$\times$254$\times$254\\
CON4 & Convolution & 64$\times$254$\times$254 & 3$\times$3 & 1 & 1  & BN & ReLU & 128$\times$128$\times$128\\
CON5 & Convolution & 128$\times$128$\times$128 & 3$\times$3 & 2 & 2  & BN & ReLU & 256$\times$128$\times$128\\
CON6 & Convolution & 256$\times$128$\times$128 & 3$\times$3 & 1 & 1  & BN & ReLU & 128$\times$128$\times$128\\
OUT & Convolution & 128$\times$128$\times$128 & 3$\times$3 & 1 & 1  & BN & ReLU & 1$\times$128$\times$128\\
\midrule
\midrule
IN & Convolution & 1$\times$128$\times$128 & 5$\times$5 & 1 & 2 &  ... & LeakyReLU & 32$\times$64$\times$64 \\
MAX2 & Max-pooling & 32$\times$64$\times$64 & 2$\times$2 & 2 & 0 & ... & ... & 32$\times$32$\times$32\\
AVG3 & Avg-pooling & 32$\times$32$\times$32 & 2$\times$2 & 2 & 0 & ... & ... & 32$\times$32$\times$32\\
CON4 & Convolution & 32$\times$32$\times$32 & 5$\times$5 & 1 & 1 & ... & LeakyReLU & 64$\times$16$\times$16\\
MAX5 & Max-pooling & 64$\times$16$\times$16 & 2$\times$2 & 2 & 0 & ... & ... & 64$\times$8$\times$8\\
AVG6 & Avg-pooling & 64$\times$8$\times$8 & 2$\times$2 & 2 & 0 & ... & ... & 64$\times$8$\times$8\\
FL7   & FlattenLayer& 64$\times$8$\times$8 & ...           &...&...& ... & ... & 1$\times$4096\\
FC8  & Fully connected & 1$\times$4096 & ... & ... & ...& ...& LeakyReLU & 1$\times$1024\\
OUT  & Fully connected & 1$\times$1024 & ... & ... & ...& ...& Sigmoid & 1$\times$1\\
\bottomrule
\label{label:G}
\end{tabular}
\end{table}

\begin{table}
\centering
\caption{The Number of Solar Magnetogram Samples of $D$}
\label{GAN_data_number}
\begin{tabular}{ccccc} 
\toprule
Solar Class & No-flare & C Class& M Class& X Class \\
\midrule

Number & 10272 & 10752 & 9120 & 9366 \\
\bottomrule
\end{tabular}
\end{table}

\begin{table}
    
    \centering
    \caption{The Number of Solar Magnetogram Samples of 6 Separate Data Sets }

	\label{table:m}
	\begin{tabular}{ccccc} 
    \toprule
        Data set & No-flare & C Class& M Class& X Class\\
    \midrule
		
		No.1 Train& 8610 & 8450 & 8200 & 8254\\
		Test & 939 & 1004 & 1033 & 948\\
	\midrule
		No.2 Train& 7971 & 8035 & 8080 & 7840\\
		Test & 1172 & 954 & 1253 & 948\\
	\midrule
		No.3 Train& 7932 & 7903 & 7847 & 7984\\
		Test & 1035 & 1253 & 955 & 948\\
	\midrule
		No.4 Train& 8120 & 8011 & 7987 & 8284\\
		Test & 1501 & 1093 & 1324 & 948\\
	\midrule
		No.5 Train& 8058 & 8040 & 8261 & 7947\\
		Test & 1001 & 1046 & 1378 & 948\\
	\midrule
		No.6 Train& 8129 & 8009 & 8187 & 7994\\
		Test & 901 & 1046 & 1124 & 948\\
    \bottomrule
	\end{tabular}
\end{table}

\section{CNN-Based Forecasting Model Within Full Solar Cycle}\label{sec:solar_cycle_cnn_model}
 Referring to the model design of Alex Net \citep{krizhevsky2012imagenet} and VGG Net \citep{simonyan2014very}, we designed a full solar cycle hybrid CNN model ($M$) for forecasting solar flare. To improve efficiency, we simplified the 4 classification problems involved in flare forecasting into 3 dichotomous problems. As shown in Table \ref{label:Mdp}, the forecasting model $M$ is composed of three sub-models, i.e., $M_{1}$, $M_{2}$ and $M_{3}$. $M_{1}$ is used to forecast M-class flares and higher levels, $M_{2}$ is used to determine whether there will be a C-class flare or not. Meanwhile, $M_{3}$ is used to identify M and X-class flare.

For $M_{1}$, we designed 5 convolutional layers and 3 fully connected layers. In the first 2 convolutional layers, we used 11$\times$11 and 5$\times$5 kernel size respectively to convolve the main features of the image. The rest of the three convolutional layers are used to convolve detailed features. For each convolutional layer, we also chose BN and ReLU as regularization and activation. In addition, we added max-pooling with 2$\times$2 kernel size to each of the last three convolutional layers. 
The feature maps obtained by convolution are multidimensional data. Therefore, it needs to be linearized before inputting to the fully connected layer. We added a layer between the convolution layer and the fully connected layer to perform linearization.
For each fully connected layer, we added a Dropout \citep{6179821} for regularization to prevent over-fitting. 
We chose softmax as a classifier to output final classification results, which is often used as the last activation function of a neural network to normalize the output of a network to a probability distribution over predicted output classes.

Both $M_{2}$ and $M_{3}$ contain 3 convolutional layers with 3$\times$3 convolution kernel to obtain detailed features and 4 fully connected layers. The parameter values of striding and padding for convolutional layers are both 1. We also added BN and ReLU for each convolutional layer. A max-pooling layer with 2$\times$2 kernel size followed each convolutional layer. For each fully connected layer, we added a Dropout for regularization to prevent over-fitting. Classification results were output finally through by the classifier softmax as well.

\begin{table*}
\centering
\caption{The structure of $M$, three models are $M_{1}$, $M_{2}$, $M_{3}$ from top to bottom}
\begin{tabular}{cccccccccc}
\toprule
Layer & Type & Channel & Kernel Size & Striding & Padding & Regularization & Activation & Pooling & Output \\
\midrule
\midrule
INPUT & Convolution & 1 & 11$\times$11   & 4 & 1 &  BN & ReLU & ... & 64 \\
CON2  & Convolution & 64 & 5$\times$5  & 1 & 2 &  BN & ReLU & ... & 64 \\
CON3  & Convolution & 64 & 3$\times$3    & 1 & 1 &  BN & ReLU & Max & 64 \\
CON4  & Convolution & 64 & 3$\times$3    & 1 & 1 &  BN & ReLU & Max & 64 \\
CON5  & Convolution & 64 & 3$\times$3    & 1 & 1 &  BN & ReLU & Max & 64 \\
FL6   & FlattenLayer& 64 & ...           &...&...& ... & ...  & ... & 576\\
FC7   & Fully Connected & 576 & ...       &...&...& Dropout&...& ... & 128\\
FC8   & Fully Connected & 128 & ...       &...&...& Dropout&...& ... & 64\\
OUT   & Fully Connected & 64 & ...       &...&...& Dropout&Softmax& ... & 2\\
\midrule
\midrule
INPUT & Convolution & 1 & 3$\times$3   & 1 & 1 &  BN & ReLU & Max & 64 \\
CON2  & Convolution & 64 & 3$\times$3 & 1 & 1 &  BN & ReLU & Max & 64 \\
CON3  & Convolution & 64 & 3$\times$3 & 1 & 1 &  BN & ReLU & Max & 64 \\
FL4   & FlattenLayer& 64 & ...           &...&...& ... & ...  & ... & 16384\\
FC5   & Fully Connected & 16384 & ...       &...&...& Dropout&...& ... & 1024  \\
FC6   & Fully Connected & 1024 & ...       &...&...& Dropout&...& ... & 512\\
FC7   & Fully Connected & 512 & ...       &...&...& Dropout&...& ... & 128\\
OUT   & Fully Connected & 128 & ...       &...&...& Dropout&Softmax& ... & 2\\
\midrule
\midrule
INPUT & Convolution & 1 & 3$\times$3   & 1 & 1 &  BN & ReLU & Max & 64 \\
CON2  & Convolution & 64 & 3$\times$3 & 1 & 1 &  BN & ReLU & Max & 64 \\
CON3  & Convolution & 64 & 3$\times$3 & 1 & 1 &  BN & ReLU & Max & 64 \\
FL4   & FlattenLayer& 64 & ...           &...&...& ... & ...  & ... & 16384\\
FC5   & Fully Connected & 16384 & ...       &...&...& Dropout&...& ... & 1024  \\
FC6   & Fully Connected & 1024 & ...       &...&...& Dropout&...& ... & 512\\
FC7   & Fully Connected & 512 & ...       &...&...& Dropout&...& ... & 128\\
OUT   & Fully Connected & 128 & ...       &...&...& Dropout&Softmax& ... & 2\\
\bottomrule
\end{tabular}
\label{label:Mdp}
\end{table*}

\begin{figure}
    \centering
    \includegraphics{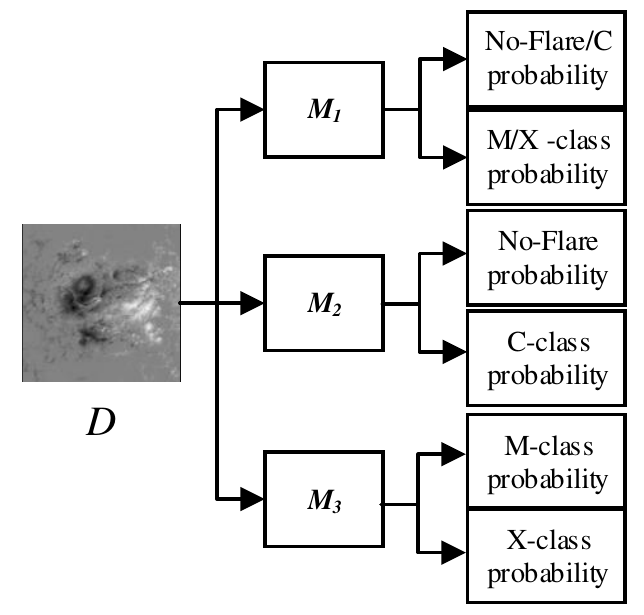}
    \caption{The training process of proposed hybrid CNN model}
    \label{fig:tr}
\end{figure}

The training process is shown in Figure \ref{fig:tr}. During training model $M$, $D$ were input to $M$, $M_{1}$ is used to output the probability of flare level $\geq$ M-class flare. The training set of $M_{2}$ is a subset extracted from the training set of $M_{1}$, including No-flare class and C-class flare samples. The role of $M_{2}$ is to forecast the probability of a C-class flare eruption. The input of $M_{3}$ is the flare data containing M-class and X-class, which is used to trained and output the probability of M-class and X-class flare.
\begin{figure}
    \centering
    \includegraphics{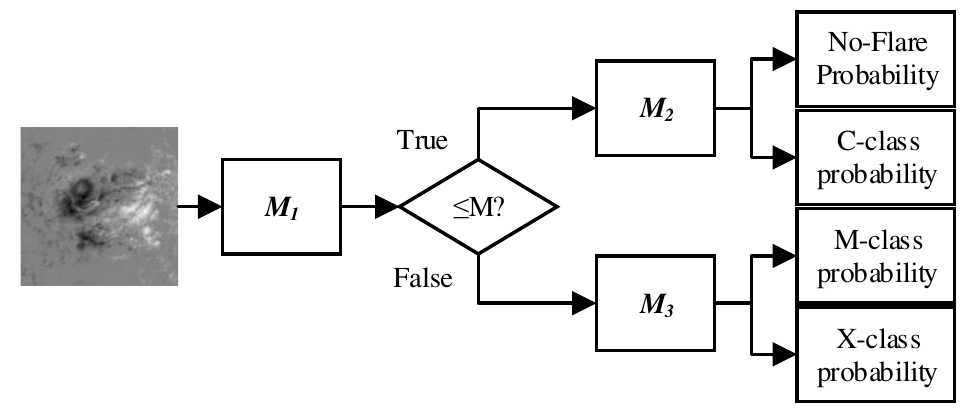}
    \caption{The training process of proposed hybrid CNN model}
    \label{fig:te}
\end{figure}

In the process of testing, as shown in Figure \ref{fig:te}, sample will be firstly input to the $M_{1}$, determine whether a $\geq$ M-class flare will erupt, if judgment is True, activate $M_{3}$, further determine whether there will be a M or X-class flare erupt within 24 hours. Otherwise, $M_{2}$ will be activated and the sample image will be input into $M_{rp2}$ to determine whether there will be a C-class flare or not within 24 hours. 

\subsection{Optimizer for Stochastic Gradient Descent}


To improve the accuracy of solar flare forecasting, we adopted the loss function based gradient descent algorithm. The core idea of the method is to minimize the loss function through training and calculate cross-entropy loss \citep{hinton2006reducing}. Pytorch provides a loss function for multi-classification problems \citep{ketkar2017introduction}.
$$
\mathop{loss}(x, \text {class})=-\log \left(\frac{\exp (x[\text {class}])}{\sum_{j} \exp (x[j])}\right)
$$
Based on it, we adopt the summation of the cross entropy as the loss function
$$
L=\sum_{n=1}^{N} \sum_{k=1}^{K}  y_{n k} \log \left(p\left(y_{n k}\right)\right)
$$
which greatly improved the efficiency of model training, where $N$ represents the number of training set samples, $K$ represents the number of the training set class, $p\left(y_{n k}\right)$ and $y_{n k}$ are the predicted output and the expected output, respectively. 

We chose Adam \citep{kingma2014adam} as the method for momentum updating. In the training process,
the samples are taken as input going through the forward propagation. Then output the results of the model, and estimate the loss function $L$. During the backward propagation, gradients of the loss function about all weights are backpropagated through the model. Gradient descent updates the weights in the convolutional layer and the fully connected layer, and the parameters of BN to minimize the loss function at the same time. The model is repeatedly trained until the loss function $L$ converges.

\begin{table}
    \centering
    \caption{The Training Parameter Settings of the $M$}
    \begin{tabular}{ccccccc}
    \toprule
         Model       & Learning Rate & Momentum &  Batch Size & Epoch\\
    \midrule
         $M_{1}$  &    0.001 & 0.5 & 16 & 50\\
         $M_{2}$  &    0.001 & 0.7 & 32 & 50\\
         $M_{3}$  &    0.001 & 0.7 & 32 & 50\\
    \bottomrule
    \end{tabular}
    \label{tab:my_label}
\end{table}

The training parameter is also crucial for achieving high predictive performance and speeding up the learning process. The parameters in our modeling are summarized in Table \ref{tab:my_label}.

\subsection{Model Validation}

Over-fitting and under-fitting are the most common problems in CNN that could lead to erroneous forecasting results. Therefore, model validation is an indispensable part of CNN.

In the process of building $M$, we track the learning performance during each epoch through training loss (Train\_loss) and validating loss (Val\_loss). Train\_loss and Val\_loss are the values of the training and validation data set output by the loss function $L$ of $M$ for each epoch.
During the learning process, Train\_loss and Val\_loss keep decreasing, indicating that the model is continuously learning. 
If Train\_loss keeps decreasing while Test\_loss tends to be constant, it may be that the model is over-fitting. We need to correct the model by adjusting the parameters of the convolutional layers and pooling layers. Moreover, regularization techniques could also be considered.
If Train\_loss tends to be constant and Test\_loss keeps decreasing, there is something wrong in the dataset collation process, and we need to re-collate the dataset. 
If both Train\_loss and Val\_loss tend to remain constant, it means that the model has encountered a bottleneck in the learning process, which can be corrected by reducing the learning rate.
However, if Train\_loss and Test\_loss keep rising simultaneously, it indicates that the network structure is unreasonable or the training hyperparameters are set improperly. The design of the model needs to be re-examined.

\begin{figure*}
\gridline{\fig{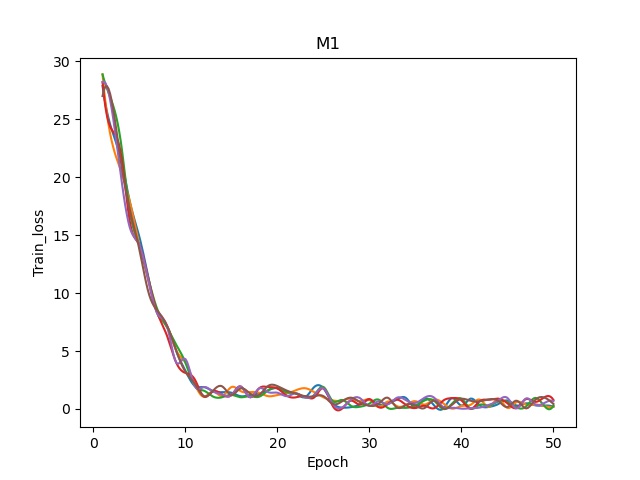}{0.34\textwidth}{(a)}
          \fig{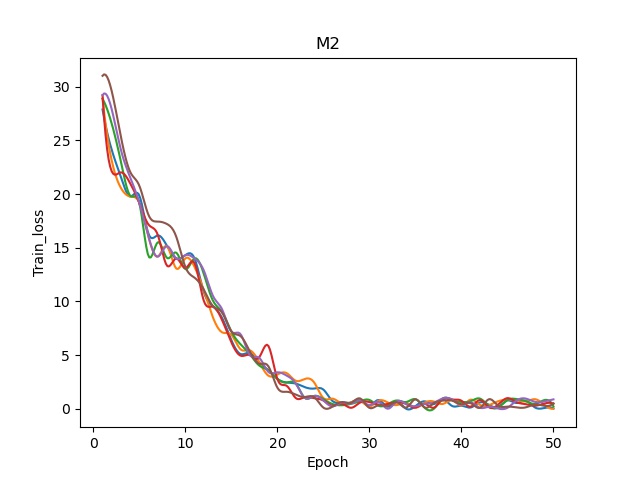}{0.34\textwidth}{(c)}
          \fig{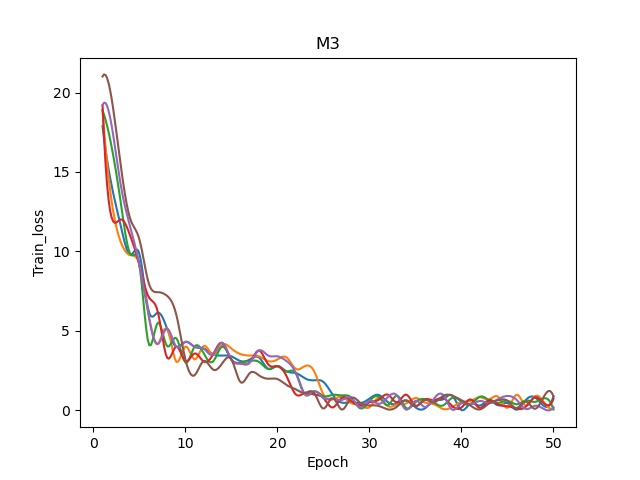}{0.34\textwidth}{(e)}
          }
\gridline{\fig{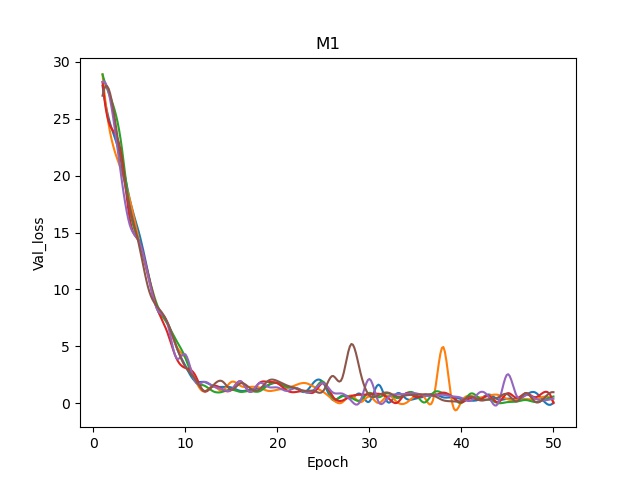}{0.34\textwidth}{(b)}
          \fig{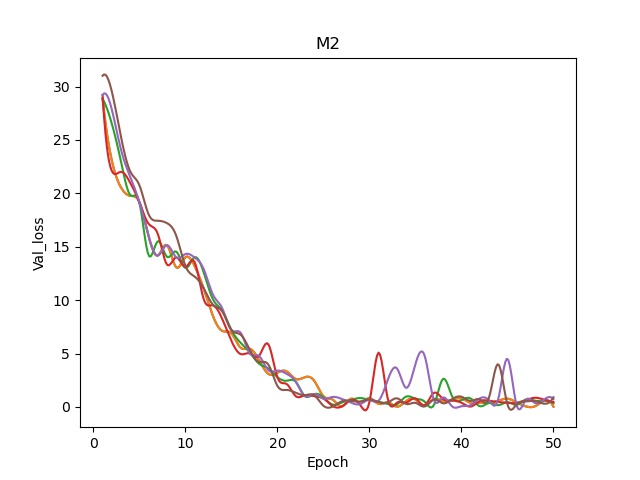}{0.34\textwidth}{(d)}
          \fig{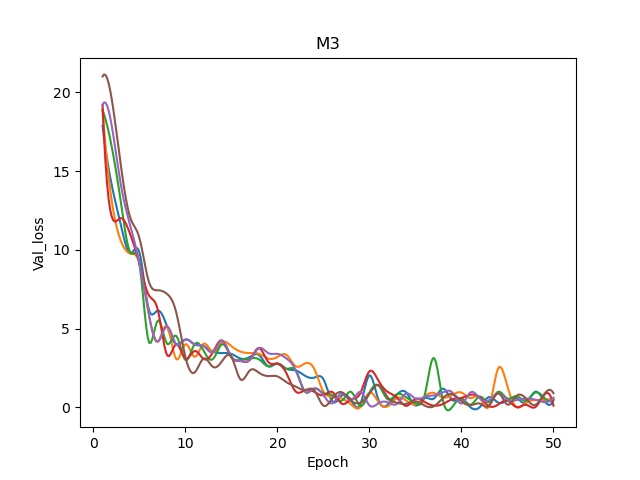}{0.34\textwidth}{(f)}
          }
\caption{Learning curves showing the result of Train\_loss and Val\_loss per epoch for the $M_{1}$, $M_{2}$ and $M_{3}$. Six different color curves show the changes of Train\_loss and Val\_loss with epochs for the model trained and tested by six separate training and testing data sets. a and b show the result of Train\_loss and Val\_loss per epoch for the model $M_{1}$, $M_{2}$ and $M_{3}$, respectively.}
\label{fig:TV}
\end{figure*}

The relationship between Train\_loss, Val\_loss and epoch in the learning process of each model is shown in Figure \ref{fig:TV}. 
In the learning process of $M_{1}$, we found that both Train\_loss and Test\_loss converge after 12 epochs.
However, the Train\_loss of $M_{2}$ appeared to converge after about 24 epochs, while Val\_loss still experienced large fluctuations from the 30 to 40 epochs, which was probably caused by samples at classification thresholds.
For $M_{3}$, there is a relatively significant difference between the six different color curves for both Train\_loss and Val\_loss. We speculate that this is due to differences within detailed features between GAN-generated samples. However, the good thing is that all started to converge after 30 epochs.
Although the relationship between Train\_loss and Val\_loss differs from the perfect state, it still shows that our models do not suffer from excessive over-fitting and under-fitting.

\subsection{Performance Metrics}

To evaluate the quality of the model, the forecast results of flare classes (No-flare, C-class, M-class and X-class) are made into a confusion matrix for evaluation. We calculated four values, i.e., True Positive (TP), True Negative (TN), False Positive (FP) and False Negative (FN), respectively. 

On the basis of statistics, we further deduced  several common metrics used for validating models.

1). Recall: the proportion of all observed True records that are forecasted to be False.
$$\mathrm{Recall}=\frac{\mathrm{TP}}{\mathrm{TP}+\mathrm{FN}}$$
2).Precision: The proportion of observed True in all records with True forecast.
$$\mathrm{Precision}=\frac{\mathrm{TP}}{\mathrm{TP}+\mathrm{FP}}$$
3).Accuracy: The proportion of correct forecasts in all records.
$$\mathrm{Accuracy}=\frac{\mathrm{TP}+\mathrm{TN}}{\mathrm{TP}+\mathrm{FN}+\mathrm{FP}+\mathrm{TN}}$$
4).Heidke skill score (HSS) \citep{heidke1926berechnung}: 
$$
\mathrm{HSS}=\frac{2[(\mathrm{TP} \times \mathrm{TN})-(\mathrm{FN} \times \mathrm{FP})]}{(\mathrm{TP}+\mathrm{FN})(\mathrm{FN}+\mathrm{TN})+(\mathrm{TP}+\mathrm{FN})(\mathrm{FP}+\mathrm{TN})}
$$
5).True skill statistics (TSS) \citep{hanssen1965relationship}:
$$
\mathrm{TSS}=\frac{\mathrm{TP}}{\mathrm{TP}+\mathrm{FN}}-\frac{\mathrm{FP}}{\mathrm{FP}+\mathrm{TN}}
$$

Among these metrics, the recall, precision and accuracy all belong to the interval of [0,1]. When the value is equal to 1, it represents the perfect forecast of the model. Therefore, the more the value is close to 1, the more mature the model is.
HSS belongs to (-$\infty$,1] and employs the whole contingency table to quantify the accuracy of achieving correct predictions, therefore it is most frequently used in flare forecasting \citep{barnes2008evaluating}. The closer the value of HSS is to 1, the more accurate the forecast of the model will be. When the value of HSS is negative, we believe that metric has no reference meaning, and we need to evaluate the model with other metrics. The interval of TSS value is [-1,1]. When TSS value is 0, this parameter value is meaningless; when TSS value is -1, it represents the worst model; when TSS value is 1, it represents the perfect model  \citep{woodcock1976evaluation}.

Since the TSS score is unbiased in the rate of class imbalance \citep{woodcock1976evaluation}, recent studies have recommended it as the primary indicator for evaluating flare forecast models \citep{2012Toward,nishizuka2018deep,zheng2019solar}. We also followed the recommendations of \cite{2012Toward}, \cite{nishizuka2018deep} and \cite{zheng2019solar}. The TSS score was used as the primary metric and the other metrics as the secondary metric in our study.

\subsection{Result}

After confirming that the models were well trained, we used the previous six separate testing sets to test the forecasting performance of $M$. The experimental results were sorted into a confusion matrix (see Table \ref{tab:cmm}) to calculate the metric score of the model and paint the receiver operating characteristic (ROC) curves to compare the performance between the 3 sub-models of $M$.
\begin{figure}
    \centering
    \includegraphics{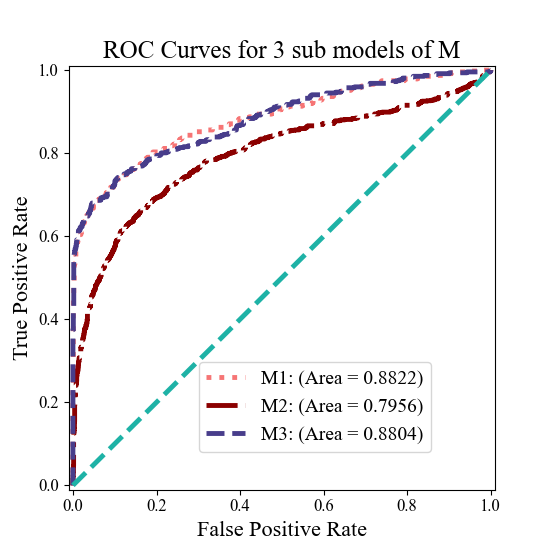}
    \caption{The ROC Curves and AUC(the Area) of 3 sub models of $M$}
    \label{fig:roc}
\end{figure}
As shown in Figure \ref{fig:roc}, the AUC of $M_{1}$, $M_{2}$, $M_{3}$ are 0.8822, 0.8804 and 0.7956 respectively. Because pooling layer may lose some valuable information and ignore the correlation between the local and the whole, We think that the result is reasonable and expected. Meanwhile, the ACU of $M_{1}$ and $M_{3}$ are better than that of $M_{2}$, which is consistent with the later experimental results. In general, $M_{1}$ and $M_{3}$ are more stable and efficient than $M_{2}$.

\begin{table}
    \centering
    \caption{The values of the confusion matrix of $M$ on Each of 6 Data Sets}

	\label{tab:cmm}
	\begin{tabular}{lcccc} 
	    \toprule
       
        Observation $\downarrow$ \\ Prediction $\rightarrow$ & No-flare  Class & C Class& M Class& X Class\\
		\midrule
		No.1:
		        No-flare& 824 & 134 & 20 & 2\\
		        C Class & 154 & 768 & 69 & 30\\
		        M Class	& 2 & 133 & 888 & 131\\
		        X Class & 1 & 66 & 69 & 747\\
		\midrule
		No.2:
		        No-flare& 817 & 126 & 27 & 1\\
		        C Class & 154 & 744 & 161 & 23\\
		        M Class	& 103 & 122 & 789 & 118\\
		        X Class & 0 & 64 & 162 & 768\\
		\midrule
		No.3:
		        No-flare& 920 & 115 & 27 & 0\\
		        C Class & 158 & 820 & 32 & 24\\
		        M Class	& 38 & 66 & 729 & 134\\
		        X Class & 5 & 24 & 257 & 752\\
		\midrule
		No.4:
		        No-flare& 713 & 135 & 4 & 13\\
		        C Class & 136 & 800 & 125 & 61\\
		        M Class	& 3 & 115 & 980 & 225\\
		        X Class & 2 & 4 & 122 & 859\\
		\midrule
		No.5:
		        No-flare& 809 & 147 & 10 & 21\\
		        C Class & 41 & 889 & 104 & 47\\
		        M Class	& 4 & 13 & 810 & 139\\
		        X Class & 0 & 5 & 117 & 751\\
		\midrule
		No.6:
		        No-flare& 613 & 117 & 4 & 0\\
		        C Class & 163 & 829 & 130 & 0\\
		        M Class	& 103 & 88 & 590 & 228\\
		        X Class & 2 & 20 & 197 & 765\\
		\bottomrule
		
	\end{tabular}
\end{table}

The means and standard deviations of these forecasting results are also presented in Table \ref{table:MResult}. $M$ has TSS scores of 0.757$\pm$0.033, 0.646$\pm$0.060, 0.653$\pm$0.078, 0.762$\pm$0.018 for No-flare, C-class, M-class, and X-class flare forecasting, respectively. Except for No-flare, the average of the TSS of C-class, M-class and X-class are significantly better than those of \cite{zheng2019solar}. The mean TSS of No-flare obtained from our model is 0.757 which is slightly lower than 0.768 of \cite{zheng2019solar}.

In addition, for M and X-class flare forecasting, the means of the TSS scores are higher than the highest values 0.653, 0.740 of \cite{zheng2019solar} and \cite{2012Toward} respectively. Moreover, the standard deviations are 0.078 and 0.018, respectively, which are much lower than 0.137 and 0.370 of \cite{zheng2019solar}. It means that our model is more robust than that of \cite{zheng2019solar}.

Meanwhile, the sub-model $M_{1}$ has a good performance for the binary classification task ($\geq$ M-class flare forecasting). The best and the worst TSS score,  calculated from the confusion matrix in Table \ref{tab:cmm},  are 0.830 and 0.718 for $\geq$ M-class flare forecasting, respectively. The scores are better than 0.800 of \cite{chen2019Identifying} , 0.539 of \cite{2012Toward},  and 0.774 (mean of TSS score) of  \cite{li2020predicting}. Moreover, the standard deviation of the sub-model $M_{1}$ is 0.056, which is also lower than 0.079 of \cite{li2020predicting}.

\begin{table}
    \caption{The Flare Prediction Results (within 24 hr) of $M$ and Comparison with Previous Studies.}
    \label{table:MResult}
    \centering

    \sisetup{add-integer-zero=false}
    \begin{tabular}{ccccccc}
        \toprule
        Metric & Model & No-flare Class & C Class & M Class & X Class & $\geq$ M Class($M_{1}$) \\
        \midrule
        & $M$ & 0.806$\pm$0.034 & 0.734$\pm$0.047 &0.746$\pm$0.072 & 0.830$\pm$0.010 & 0.890$\pm$0.041\\
        & \cite{chen2019Identifying} & ... & ... & ... & ... & 0.930\\
        Recall & \cite{zheng2019solar} & 0.869$\pm$0.034 & 0.671$\pm$0.059 & 0.617$\pm$0.148 & 0.594$\pm$0.394 & ...\\
        & \cite{li2020predicting} & ... & ... & ... & ... & 0.817$\pm$0.084\\
        & \cite{2012Toward} & ... & 0.737 & 0.693 & 0.859 & 0.704\\
        \midrule
        & $M$ & 0.849$\pm$0.012 & 0.744$\pm$0.044 & 0.740$\pm$0.031 & 0.780$\pm$0.050 & 0.907$\pm$0.029\\
        & \cite{chen2019Identifying} & ... & ... & ... & ... & 0.870\\
        Precision & \cite{zheng2019solar} & 0.793$\pm$0.054 & 0.670$\pm$0.079 & 0.699$\pm$0.087 & 0.562$\pm$0.383 & ...\\
        & \cite{li2020predicting} & ... & ... & ... & ... & 0.889$\pm$0.056\\
        & \cite{2012Toward} & ... & ... & 0.136 & 0.029 & 0.146\\
        \midrule
        & $M$ &0.913$\pm$0.008 & 0.865$\pm$0.023 & 0.864$\pm$0.024 & 0.909$\pm$0.014 & 0.886$\pm$0.029\\
        & \cite{chen2019Identifying} & ... & ... & ... & ... & ...\\
        Accuracy & \cite{zheng2019solar} & 0.891$\pm$0.018 & 0.812$\pm$0.029 & 0.849$\pm$0.034 & 0.933$\pm$0.041 & ...\\
        & \cite{li2020predicting} & ... & ... & ... & ... & 0.891$\pm$0.024\\
        & \cite{2012Toward} & ... & 0.711 & 0.829 & 0.881 & 0.830\\
        \midrule
        & $M$ & 0.751$\pm$0.038 & 0.644$\pm$0.062 & 0.655$\pm$0.085 & 0.770$\pm$0.013 & 0.764$\pm$0.071\\
        & \cite{chen2019Identifying} & ... & ... & ... & ... & 0.810\\
        HSS & \cite{zheng2019solar} & 0.747$\pm$0.037 & 0.535$\pm$0.061 & 0.551$\pm$0.120 & 0.539$\pm$0.366 & ...\\
        & \cite{li2020predicting} & ... & ... & ... & ... & 0.759$\pm$0.071\\
        & \cite{2012Toward} & ... & ... & 0.136 & 0.029 & 0.190\\
        \midrule
        & $M$ & 0.757$\pm$0.033 & 0.646$\pm$0.060 & 0.653$\pm$0.078 & 0.762$\pm$0.018 & 0.774$\pm$0.056\\
        & \cite{chen2019Identifying} & ... & ... & ... & ... & 0.800\\
        TSS & \cite{zheng2019solar} & 0.768$\pm$0.028 & 0.538$\pm$0.059 & 0.534$\pm$0.137 & 0.552$\pm$0.370 & ...\\
        & \cite{li2020predicting} & ... & ... & ... & ... & 0.749$\pm$0.079\\
        & \cite{2012Toward} & ... & 0.443 & 0.526 & 0.740 & 0.539\\
        \midrule
        & \cite{chen2019Identifying} & ... & ... & ... & ... & -3.3\% \\
        TSS Rate & \cite{zheng2019solar} & -1.4\% & 20.1\% & 22.3\% & 38.0\% & ...\\
        Improved & \cite{li2020predicting} & ... & ... & ... & ... & 3.3\%\\
        & \cite{2012Toward} & ... & 45.8\% & 24.1\% & 3.0\% & 43.6\%\\

        \bottomrule

    \end{tabular}
\end{table}

\section{Fine-grained Hybrid CNN model}\label{sec:fine-grained_cnn_model}
A solar cycle consists of rising and declining periods, and there is some difference in the intensity of solar activity between the rising and declining periods. After modeling  solar flare forecasting within a full-cycle, we realized that this difference might have some implications for model building. This section further investigates the independent modeling of the rising and declining phases separately and evaluates their performance.

\subsection{Data Set Preparation}

We continued to use the dataset $D$ on which the model $M$ was built. We divide the dataset $D$ into $D_{rp}$ and $D_{dp}$.
The $D_{rp}$ is the subset of the rising phase ($T_{rp}$) in the solar cycle 24 from May 2010 to April 2014. $D_{dp}$ is the subset of the declining phase ($T_{dp}$) from May 2014 to November 2019.



    

		

\subsection{Modeling For Rising and Declining Phases}

Two hybrid CNN models ($M_{rp}$ and $M_{dp}$) are built for rising and declining phases, respectively. $M_{dp}$ is the same as $M$. However, there is a certain difference between $M_{rp}$ and $M$.
$M_{rp}$ also contains 3 CNN model, i.e.,  $M_{}rp1$, $M_{rp2}$ and $M_{rp3}$. However, we modified the convolution kernels of the first two convolutional layers to 15$\times$15 and 11$\times$11 compared to $M_{1}$ for $M_{rp1}$. As for $M_{rp2}$ and $M_{rp3}$, we added a layer after CON3 (see Table \ref{label:Mdp}) which is the same as CON3, and removed a fully connected layer. The specific modeling and tuning process is similar to the previous modeling and tuning of $M$.

It is necessary to note that there is not much objective basis for such a model change. We finally built this model through extensive tuning and testing. The modification of the model may be caused by the different characteristics of the magnetic fields in the rising and declining phases.
Solar flare releases its energy previously stored in the magnetic field, which is a rapid transformation process of magnetic energy of the solar active regions into the kinetic energy of plasma flow, particle, radiation, and heat. As the duration of the rising phase is shorter than the declining phase of a solar cycle, that is to say, the physical process of flare energy storage and dissipation in the two phases may exhibit different spatio-temporal behaviors \citep{hudson2011global}.




\subsection{Results of $M_{rp}$ and $M_{dp}$}
Precisely as in the previous sections, we evaluated the performance of $M_{rp}$ and $M_{dp}$, still mainly on the TSS.

First of all, we input the sub testing sets of $D$ into the $M_{rp}$ and $M_{dp}$, respectively, and make the output results of the model into  Table \ref{table:MrpdpResult} and Table \ref{table:MrpdpResult2} according to the confusion matrix. Then calculate the mean and standard deviation for each solar flare class forecast, which is recorded with $M$ in Table \ref{tab:t3}.

According to Table \ref{tab:t3}, all model evaluation metrics of $M_{rp}$ are better than $M$ in the flare forecast of No-flare, C, M and X-class. The TSS score of $M_{rp}$, they are up to 0.802$\pm$0.013, 0.707$\pm$0.018, 0.770$\pm$0.042 and 0.862$\pm$0.064, respectively. Meanwhile, the TSS score of $M_{dp}$ for No-flare, C-class, M-class, and X-class were better than $M$, which were 0.768$\pm$0.036, 0.663$\pm$0.032, 0.728$\pm$0.039 and 0.855$\pm$0.007. However, for the No-flare and C-class flares forecast, the difference of mean between $M$ and $M_{dp}$ is only 0.012 and 0.017. We realized this is a reasonable result caused by the same model structure. However, in the forecasting of M and X-class, the TSS score has greatly improved by 17.9\%, 13.1\%. Combined with the forecast result of $M_{rp}$, it proved that investigating the modeling for the rising phase and the declining phase is an effective method to improve the accuracy of forecasting solar flare.

\begin{table}
    \centering
    \caption{Values of confusion matrix for testing  $M_{rp}$ on each of 6 Data Sets}

	\label{table:MrpdpResult}
	\begin{tabular}{lcccc} 
	    \toprule
            Observation$\downarrow$ \\ Prediction$\rightarrow$    

& No-flare & C Class & M Class & X Class\\
		\midrule
		No.1:
		        No-flare& 782 & 134 & 0 & 7\\
		        C Class & 133 & 805 & 88 & 16\\
		        M Class	& 18 & 60 & 887 & 31\\
		        X Class & 6 & 5 & 58 & 894\\
		\midrule
		No.2:
		        No-flare& 965 & 59 & 1 & 1\\
		        C Class & 121 & 747 & 58 & 15\\
		        M Class	& 82 & 91 & 853 & 34\\
		        X Class & 4 & 57 & 104 & 898\\
		\midrule
		No.3:
		        No-flare& 903 & 168 & 1 & 0\\
		        C Class & 116 & 927 & 33 & 23\\
		        M Class	& 14 & 108 & 759 & 129\\
		        X Class & 2 & 50 & 162 & 796\\
		\midrule
		No.4:
		        No-flare& 1337 & 44 & 3 & 5\\
		        C Class & 109 & 777 & 176 & 21\\
		        M Class	& 28 & 248 & 1004 & 32\\
		        X Class & 27 & 24 & 141 & 890\\
		\midrule
		No.5:
		        No-flare& 857 & 163 & 67 & 2\\
		        C Class & 104 & 825 & 183 & 9\\
		        M Class	& 20 & 39 & 986 & 28\\
		        X Class & 20 & 19 & 142 & 909\\
		\midrule
		No.6:
		        No-flare& 766 & 140 & 0 & 0\\
		        C Class & 88 & 854 & 81 & 3\\
		        M Class	& 26 & 42 & 868 & 45\\
		        X Class & 21 & 10 & 175 & 900\\
		\bottomrule

	\end{tabular}
\end{table}

\begin{table}
    \centering
    \caption{Values of confusion matrix for testing  $M_{dp}$ on each of 6 Data Sets}

	\label{table:MrpdpResult2}
	\begin{tabular}{lcccc} 
	    \toprule
            Observation$\downarrow$ \\ Prediction$\rightarrow$    

& No-flare & C Class & M Class & X Class\\
	    \midrule
    No.1:
          No-flare& 849 & 179 & 0 & 0\\
          C Class & 174 & 750 & 84 & 41\\
          M Class & 63 & 33 & 796 & 71\\
          X Class & 0 & 0 & 68 & 836\\
        \midrule
    No.2:
          No-flare& 887 & 115 & 20 & 0\\
          C Class & 132 & 800 & 149 & 42\\
          M Class & 60 & 192 & 916 & 67\\
              X Class & 0 & 2 & 61 & 801\\
        \midrule
    No.3:
          No-flare& 493 & 32 & 0 & 0\\
          C Class & 91 & 492 & 55 & 0\\
          M Class & 10 & 92 & 417 & 30\\
          X Class & 0 & 0 & 52 & 210\\
        \midrule
    No.4:
          No-flare& 1056 & 40 & 8 & 33\\
          C Class & 298 & 961 & 112 & 9\\
          M Class & 126 & 79 & 873 & 38\\
          X Class & 21 & 13 & 85 & 830\\
        \midrule
    No.5:
          No-flare& 909 & 19 & 32 & 0\\
          C Class & 286 & 968 & 206 & 0\\
          M Class & 33 & 143 & 915 & 38\\
          X Class & 19 & 6 & 164 & 872\\
    \midrule
    No.6:
          No-flare& 1058 & 79 & 9 & 0\\
          C Class & 287 & 861 & 90 & 2\\
          M Class & 131 & 113 & 1082 & 7\\
          X Class & 25 & 40 & 97 & 901\\
  \bottomrule
	\end{tabular}
\end{table}

\begin{table}

\centering
\sisetup{add-integer-zero=false}
\caption{The flare forecasting results (within 24 hours) of $M_{rp}$ and $M_{dp}$}
\label{tab:t3}
\begin{tabular}{cccccc}
\toprule
Metric & Model & No-flare Class & C Class & M Class & X Class  \\
\midrule
 & $M$ & 0.806$\pm$0.034 & 0.734$\pm$0.047 &0.746$\pm$0.072 & 0.830$\pm$0.010\\
 Recall & $M_{rp}$ & 0.842$\pm$0.021 & 0.774$\pm$0.026 & 0.830$\pm$0.027 & 0.910$\pm$0.050\\
 & $M_{dp}$ & 0.811$\pm$0.021 & 0.766$\pm$0.033 & 0.811$\pm$0.020 & 0.879$\pm$0.003\\
 \midrule
 & $M$ & 0.849$\pm$0.012 & 0.744$\pm$0.044 & 0.740$\pm$0.031 & 0.780$\pm$0.050\\
 Precision & $M_{rp}$ & 0.876$\pm$0.045 & 0.802$\pm$0.030 & 0.815$\pm$0.057 & 0.853$\pm$0.058\\
 & $M_{dp}$ & 0.877$\pm$0.047 & 0.732$\pm$0.027 & 0.775$\pm$0.037 &0.884$\pm$0.059\\
 \midrule
 & $M$ &0.913$\pm$0.008 & 0.865$\pm$0.023 & 0.864$\pm$0.024 & 0.909$\pm$0.014\\
 Accuracy & $M_{rp}$ & 0.929$\pm$0.004 & 0.890$\pm$0.008 & 0.912$\pm$0.017 &0.942$\pm$0.023\\
 & $M_{dp}$ & 0.916$\pm$0.016 & 0.861$\pm$0.008 & 0.889$\pm$0.021 & 0.957$\pm$0.002\\
 \midrule
 & $M$ & 0.751$\pm$0.038 & 0.644$\pm$0.062 & 0.655$\pm$0.085 & 0.770$\pm$0.013\\
 HSS & $M_{rp}$ & 0.796$\pm$0.019 & 0.702$\pm$0.028 & 0.772$\pm$0.037 & 0.871$\pm$0.064\\
 & $M_{dp}$ & 0.757$\pm$0.033 & 0.668$\pm$0.035 & 0.734$\pm$0.036 & 0.852$\pm$0.002\\
 \midrule
 & $M$ & 0.757$\pm$0.033 & 0.646$\pm$0.060 & 0.653$\pm$0.078 & 0.762$\pm$0.018\\
 TSS & $M_{rp}$ & 0.802$\pm$0.013 & 0.707$\pm$0.018 & 0.770$\pm$0.042 & 0.862$\pm$0.064\\
 & $M_{dp}$ & 0.768$\pm$0.036 & 0.663$\pm$0.032 & 0.728$\pm$0.039 & 0.855$\pm$0.007\\
 \midrule
 TSS (Mean) & $M_{rp}$ & 5.9\% & 9.4\% & 17.9\% & 13.1\% \\
 improved over M & $M_{dp}$ & 1.5\% & 2.6\% & 11.5\% & 12.2\% \\
 
\bottomrule
\end{tabular}
\end{table}

\section{Discussions}\label{sec:discussion}

\subsection{Augmented Data Validation}
The number of different samples required in machine learning is balanced. The trustworthiness of the data added by GAN is a basis for the whole research work.

To valiadate the samples augmented by the GAN, we use the X-class samples augmented by GAN as the training data and the original X-class samples as the testing data for model $M$.


\begin{table}
    \centering
    \caption{The confusion matrix of $M$ with and without X-class sample data augmented from left to right}
    \begin{tabular}{c c c c | c c c c }
        \toprule
            Observation$\downarrow$ & &  & & Observation$\downarrow$ \\ Prediction$\rightarrow$    
        & Others & X-Class & &   Prediction$\rightarrow$ &Others & X-Class\\
        \midrule
        Others & 887 & 167 & &  Others & 835 & 711\\
        \midrule
        X-Class & 132 & 801 & & X-Class & 126 & 136\\
        \bottomrule
    \end{tabular}
    \label{tab:with}
\end{table}


Firstly, We chose two training sets, i.e., the generated X samples and the original X samples, to train $M$ and then use the rest of the original sample to test. 801 out of 968 X-class samples were identified with X-class sample data augmented, 711 out of 846 samples were misidentified without X-class sample data augmented (see Table \ref{tab:with}). We found that the result of $M$ is very unsatisfactory due to too few original samples of X-class, but it has been apparently improved after using GAN to augment the number of X-class samples. Therefore, we realized that: 1) Too few samples of X-class directly impact the quality of the model. 2) The X-class samples generated by GAN are reliable and can be used as the training set of $M$.

\subsection{The Implications of Three Models}
In the study, we built a hybrid CNN model ($M$) for forecasting solar flares within 24 hours in a full solar cycle. Then we built $M_{rp}$ and $M_{dp}$ for the rising and declining phase by using the samples in different phases. All three models achieved reliable forecasting results for their periods. 

However, from the modeling point of view, the model structure of $M$ and $M_{dp}$ is basically the same. $M_{rp}$ increases the number of layers and modifies the kernel size on top of $M$, which raises a question worth discussing whether $M_{rp}$ is a more accurate predictor of flares occurring over the full solar cycle.
We have tried to use $M_{rp}$ to forecast solar flares in $T$, the means of TSS score are not ideal, which are 0.470, 0.312, 0.549, 0.642 for No-flare, C-class. M-class, X-class flare forecasting, respectively.

Therefore, we believe that the three models have their own forecasting significance in their own periods. When we cannot identify the period which the sample belongs to, we can use $M$ to forecast. Conversely, we use the corresponding model to forecast.

\section{Conclusions And Future work}

In the study, we first augmented the X-class samples using the GAN and then divided the augmented X samples and real samples into the training and testing sets, respectively. Then balanced the proportion of samples in each category by using other methods such as under-sampling. After organizing the dataset, we proposed a hybrid convolutional neural network model ($M$) for forecasting flare eruption in a solar cycle.
Basing on this model, we further investigated the effects of the rising and declining phases for flare forecasting. Two-hybrid CNN models, i.e., $M_{rp}$ and $M_{dp}$, were presented to forecast solar flare eruption in the rising phase and declining phase of the solar cycle, respectively. 

Overall, all three flare forecasting models developed in this study have excellent accuracy and stability. Significantly, the accuracy of the two models for the rising and declining phases is more advantageous, which provides a more reliable means to carry out space weather forecasting in the later stage.

\begin{acknowledgments}
This work is supported by the National SKA Program of China (2020SKA0110300), the Joint Research Fund in Astronomy (U1831204, U1931141) under cooperative agreement between the National Natural Science Foundation of China (NSFC) and the Chinese Academy of Sciences (CAS), the Funds for International Cooperation and Exchange of the National Natural Science Foundation of China (11961141001). 
This work is also supported by Astronomical Big Data Joint Research Center, co-founded by National Astronomical Observatories, Chinese Academy of Sciences and Alibaba Cloud. 

Thanks to SDO and NASA for the data and data information. 

The authors gratefully acknowledge the helpful comments and suggestions of the reviewers.
\end{acknowledgments}


\bibliography{sample631}{}
\bibliographystyle{aasjournal}



\end{document}